\documentclass[11pt]{article}

\usepackage[a4paper,margin=1in]{geometry}
\usepackage[parfill]{parskip}
\usepackage{amsmath,amssymb,amsthm}
\usepackage{graphicx}
\usepackage{multirow}
\usepackage{array}
\usepackage{moreverb}
\usepackage{authblk}
\usepackage{hyperref}
\usepackage{graphicx}
\usepackage{booktabs}
\usepackage{caption}
\usepackage{tabularx}
\usepackage{threeparttable}
\usepackage{placeins}
\usepackage[superscript]{cite}

\newcommand\BibTeX{{\rmfamily B\kern-.05em \textsc{i\kern-.025em b}\kern-.08em
T\kern-.1667em\lower.7ex\hbox{E}\kern-.125emX}}

\title{Assessing Covariate-Adjusted Risk Differences in Small-Sample Clinical Trials}

\author[1]{Martin Schnuerch}
\author[2]{Alex Ocampo}
\author[3]{Klaus Kähler Holst}
\author[1]{Christian Stock}

\affil[1]{Global Biostatistics and Data Sciences, Boehringer Ingelheim Pharma GmbH \& Co. KG, Germany}
\affil[2]{Statistical Methodology, F. Hoffmann-La Roche Ltd, Basel, Switzerland}
\affil[3]{Biostatistics Methods, Novo Nordisk, S{\o}borg, Denmark}

\date{} 

\begin{document}

\maketitle

\begin{abstract}
\noindent Binary endpoints are common in clinical trials and conditional odds ratios have traditionally been used to assess treatment effects. However, the interpretation of odds ratios is difficult, they are non-collapsible, and conditional odds-ratios obtained from regression models additionally rely on modeling assumptions in order to be a relevant overall summary measure for the trial. As an alternative, risk differences have gained increasing prominence as a more interpretable, clinically meaningful and assumption-lean measure of treatment effects. This shift has also been motivated by new regulatory guidance, which emphasizes the relevance of marginal estimands and encourages covariate adjustment. Yet, covariate-adjusted inference for risk differences, particularly in smaller samples, has methodological subtleties and lacks well-established best practices. We conduct a simulation study comparing methods for estimating and testing risk differences in small-sample (N$\,\leq\,$150) randomized clinical trials with prognostic categorical baseline covariates, focusing on exact unconditional tests, Mantel-Haenszel methods, and $g$-computation (standardization) approaches. We find that several $g$-computation approaches exhibit inflated Type I error in very small samples when standard Wald-type inference is applied, whereas robust or penalized variants improve error control at the expense of power. Classical methods such as the Mantel-Haenszel and Suissa-Shuster tests remain robust but may forgo efficiency gains from covariate adjustment. Overall, our results suggest that misalignment between estimand and variance estimation may contribute to the Type I error inflation, beyond the impact of small sample size alone. Based on these results, we provide practical recommendations to guide method selection that align the estimand, variance estimation, and inferential target.
\end{abstract}


\maketitle

\section{Introduction}
\label{sec:introduction}

Binary endpoints are widely used in randomized clinical trials, and a common objective is to compare event probabilities between treatment arms. Examples include overall response in oncology and clinical remission in ulcerative colitis trials. 
For such comparisons, odds ratios, particularly \textit{conditional} odds ratios, have long been a preferred summary measure. Their popularity stems in part from the fact that they can be obtained straightforwardly from multivariable logistic regression models, in which regression coefficients represent conditional log odds ratios. Despite their frequent use, odds ratios have well-recognized limitations and have been subject to criticism with respect to interpretability, clinical meaningfulness, and underlying assumptions.\cite{Sackett1996,Kim2006,Newcombe2013} Odds ratios are difficult to interpret, because they do not directly quantify differences or ratios of event probabilities. They are often seen as an approximation of the relative risks of an event, which is only valid when the outcome is relatively uncommon, and they always exaggerate the strength of the relative risk. Moreover, odds ratios are non-collapsible, i.e.\ even when conditional odds ratios are constant across covariate strata, the corresponding marginal odds ratio generally differs and is trending closer to the null in the presence of any prognostic factor.\cite{Daniel2021}

Recent guidance from the US Food and Drug Administration (FDA) on adjustment for covariates in randomized clinical trials highlighted the importance of distinguishing between conditional and marginal treatment estimands.\cite{FDA2023Covariates,VanLancker2024a,Wei2024} In this context, the terms \textit{marginal}, \textit{unconditional}, and \textit{population-average} refer to the same population-level treatment effect, defined by averaging over the distribution of baseline covariates. While marginal and conditional estimands coincide in linear models without treatment-by-covariate interactions, this equivalence generally breaks down in nonlinear models, such as logistic regression for binary endpoints due to non-collapsibility. Although conditional effects are, in principle, most relevant from an individual patient perspective in a clinical setting, they depend on a correctly specified model and can therefore generally not be estimated reliably in a clinical trial. From a regulatory and public-health perspective, marginal estimands align most directly with the objective of quantifying average treatment benefit in the target population. It is also the estimand targeted by an unadjusted analysis. The FDA guidance favors approaches that provide valid inference under approximately the same minimal statistical assumptions as unadjusted analyses, which themselves are considered acceptable. Conditional estimands, in the FDA guidance on covariate adjustment, are framed primarily as supportive, exploratory, or subgroup-specific targets rather than default choices for primary inference. The focus on population-level summaries is consistent with the ICH E9(R1) estimand framework, which requires explicit prespecification of the population-level summary measure aligned with the clinical question of interest before choosing the estimator and estimation method.\cite{ICHE9R12019} 

As a consequence of the limitations of (conditional) odds ratios and stimulated by the recent FDA guidance, there has been increasing interest in assessing risk differences. Unlike odds ratios, they directly quantify absolute differences in event proportions and convey treatment impact on an absolute scale. Further, they are collapsible and avoid interpretational ambiguities. Despite conceptual simplicity, estimation and inference for risk differences are not necessarily straightforward in the presence of covariates. Statistical methods to assess covariate-adjusted risk differences in clinical trials include Cochran-Mantel-Haenszel approaches (limited to categorical covariates),\cite{Mantel1959, Mantel1980, Agresti2013} and various implementations of $g$-computation (standardization),\cite{Robins1987, Liu2024} and inverse probability of treatment weighting.\cite{Glynn2010, Williamson2014} In addition, regression-based approaches that directly model the risk difference by combining a model for the risk difference with an odds-product nuisance parameterization\cite{Richardson2017}, and semiparametric methods such as targeted learning\cite{vanderLaan2011} have been proposed to enable flexible and potentially assumption-lean estimation. These methods may differ with respect to targeted estimands, reliance on modeling assumptions, computational complexity, and finite-sample behavior. As an alternative, one may forgo potential efficiency gains from covariate adjustment and apply (unconditional) unadjusted statistical tests.\cite{Korn2024, Suissa1985} 

The challenges in assessing covariate-adjusted risk differences warrant closer scrutiny in small-sample settings. Trials with total sample sizes of N$\,\leq\,$150 are common in phase~II development, and, more broadly, in rare diseases and pediatric indications, where feasibility constraints limit enrollment. First, many applied inferential procedures are asymptotically valid but provide no guarantees in finite samples.\cite{Bannick2026} In small trials, adjusted risk difference estimators may exhibit bias, non-normal sampling distributions, variance underestimation, miscalibrated $p$-values, and confidence interval undercoverage. For example, parametric variance estimators for $g$-computation based on the delta method, as well as variance estimators for the Mantel-Haenszel risk difference, are derived under large-sample theory assuming asymptotic normality, which in turn relies on large-sample properties. Similarly, the Cochran-Mantel-Haenszel test is based on a statistic that is asymptotically $\chi^2$-distributed and therefore requires a minimum amount of information per stratum to perform as intended. When sample sizes are small, events are rare, or stratification leads to very small strata, these conditions may not be met. In such settings, degrees-of-freedom penalties and numerical instability could offset potential gains from stratification. Second, $g$-computation approaches for binary endpoints depend on a fitted multivariable logistic regression working model, which may suffer from instability, separation, or non-convergence in small samples. Under these circumstances, the comparative advantage of covariate-adjusted approaches over simpler unadjusted methods remains uncertain.

To address these concerns, we present a simulation-based comparative evaluation of a range of methods for estimating and testing risk differences in randomized clinical trials, particularly exact unconditional testing, Mantel-Haenszel methods, and $g$-computation (standardization) approaches. Our focus is on small-sample settings, where misalignment between the target estimand, point estimator, and variance estimator is most likely to be revealed, and that have been studied only to a limited extent. Our intent is to inform statistical practice by identifying approaches that target the intended estimand, exhibit robust inferential performance in small samples, and strike a balance between efficiency, robustness, and interpretability.

\section{Hypothetical trial example and potential estimands}
\label{sec:hypotheticaltrial}

Consider a hypothetical two-arm randomized controlled trial with parallel groups, comparing an investigational treatment ($A=1$) with a control ($A=0$). Let $Y \in \{0,1\}$ denote a binary primary endpoint indicating treatment response, where $Y=1$ corresponds to response. Assume further that two binary baseline covariates $X=(X_1,X_2)$ are measured and considered prognostically relevant, for example sex ($X_1 \in \{\text{male},\text{female}\}$) and disease severity ($X_2 \in \{\text{mild},\text{severe}\}$). These covariates may be used as stratification factors in the randomization, and adjustment for them in the statistical analysis is generally advisable to improve statistical efficiency.

A critical step in selecting an appropriate statistical method for analysis is the clear definition of the estimand. An estimand provides a precise and comprehensive specification of what the trial aims to estimate, encompassing the treatment under investigation, the target population, the endpoint, the handling of intercurrent events, and the summary measure used to quantify the treatment effect.\cite{ICHE9R12019,Keene2023,RemiroAzocar2025} Each of these components is important; however, in this work we focus on the choice of the summary measure and therefore use the term \emph{estimand} in this broader sense.

To formalize different estimands of interest, we introduce potential outcomes notation. Let $Y^{(A)}$ denote the \emph{potential outcome} for an individual under treatment assignment $A$. Within the potential outcomes framework,\cite{Hernan2024} the treatment effect is defined as a contrast between the outcomes that would be observed for the same individual under each treatment condition. In a two-arm randomized controlled trial, only one of the two potential outcomes is observed for each individual, while the counterfactual outcome remains unobserved. 

Utilizing the potential outcomes framework, multiple estimands of interest can be defined, with a central consideration being how the treatment effect is characterized in relation to baseline covariates. A key distinction is between \emph{marginal} and \emph{conditional} estimands. Marginal estimands are defined by averaging over the distribution of baseline covariates in the target population, whereas conditional estimands describe treatment effects at fixed covariate values. While conditional estimands are well-defined in general, their reliable estimation may be challenging in settings with high-dimensional or continuous covariates. In the present work, this issue is mitigated by the fully discrete covariate structure.

We focus on the risk difference (RD) as summary measure of the treatment effect. Accordingly, all estimands are formulated as differences in response probabilities under treatment and control. Importantly, all estimands discussed herein could equivalently be defined using other contrasts (e.g., risk ratios or odds ratios). Therefore, the distinctions considered here pertain to the role of covariate adjustment rather than the choice of summary measure.

We discuss three estimands that are particularly relevant in the context of clinical trials, following Magirr et~al.\ \cite{Magirr2025}: the \textit{marginal treatment effect}, MTE, also referred to as the (population-)average treatment effect, the \textit{conditional treatment effect} at a specific covariate value $x$, CTE($x$), and the \textit{conditional population-average treatment effect}, CPATE, i.e. the average CTE($x$) over the empirical covariate distribution observed in the trial. Table \ref{tab:estimands} provides the formal definitions and interpretation of these three estimands. 

\begin{table}[ht]
\centering
\footnotesize
\caption{Summary of estimands expressed on the risk difference scale.}
\label{tab:estimands}
\renewcommand{\arraystretch}{1.3}
\begin{threeparttable}
\begin{tabularx}{\textwidth}{l l X}
\hline
\textbf{Estimand} & \textbf{Definition} & \textbf{Interpretation} \\
\hline
MTE 
& $\Pr(Y^{(1)}=1) - \Pr(Y^{(0)}=1)$ 
& \emph{What is the average treatment effect if all individuals in the population were treated versus untreated?} \\

CTE$(x)$ 
& $\Pr(Y^{(1)}=1 \mid X=x) - \Pr(Y^{(0)}=1 \mid X=x)$ 
& \emph{What is the average treatment effect for individuals with covariate value $x$?} \\

CPATE 
& $\frac{1}{N} \sum_{i=1}^N \left\{\Pr\!\left(Y^{(1)}=1 \mid X = x_i\right) -\Pr\!\left(Y^{(0)}=1 \mid X = x_i\right)\right\}$ 
& \emph{What is the average of covariate-specific treatment effects across the population as observed in the trial?} \\
\hline
\end{tabularx}

\begin{tablenotes}
\footnotesize
\item Abbreviations: MTE = marginal treatment effect; CTE = conditional treatment effect; CPATE = conditional population-average treatment effect.
\end{tablenotes}
\end{threeparttable}
\end{table}

This selection of estimands does not constitute an exhaustive taxonomy but represents those most relevant for the present work. While a decision for or against a CTE may often appear straightforward guided by the clinical objective, the choice between MTE and CPATE is more subtle. Magirr et al.\cite{Magirr2025} argue that there is no universally correct choice between these estimands; rather, the estimand of interest should be transparently prespecified, and the statistical method should be selected to target that estimand. They further argue that the CPATE, although defined through covariate-specific effects, may reasonably be interpreted as a marginal estimand because it averages these effects over the covariate distribution observed in the trial. We concur with this perspective. 
Indeed, from an estimation perspective, both estimands correspond to marginal population-level contrasts of potential outcomes and may therefore be algebraically identical at the level of point estimation under standardization-based approaches. Their distinction instead lies in the inferential framework: the CPATE conditions on the observed covariate distribution, whereas the MTE treats this distribution as random and accounts for its variability. Consequently, the choice between these estimands is reflected not in the point estimator but in the variance estimator, confidence intervals, and hypothesis tests, which must be aligned with the chosen estimand.
Rather than advocating for a single default estimand, we emphasize the importance of acknowledging the diversity of clinically meaningful estimands, clearly communicating the chosen definition, and ensuring alignment between the estimand, the estimator, and the inferential procedure. 

\section{Statistical Methods}
\label{sec:statmethods}

In this section, we describe the statistical methods considered in this study. The selected methods are intended to represent a range of approaches applicable to the hypothetical trial setting described above.
For each method, we identify the targeted estimand and summarize the key assumptions underlying estimation and inference. Table~\ref{tab:methodsOverview} provides an overview of all evaluated methods, including their form of covariate adjustment, inferential approach, and targeted estimand. Appendix \ref{app:software} gives an overview of software packages to implement the methods.

\begin{table}[ht]
\centering
\small
\caption{Overview of statistical methods evaluated in simulation.}
\label{tab:methodsOverview}
\begin{tabularx}{\textwidth}{l l l l}
\toprule
\textbf{Method} &
\textbf{Covariate adjustment} &
\textbf{Inference} &
\textbf{Estimand} \\
\midrule
\multicolumn{4}{l}{\textit{Unadjusted methods}} \\
Suissa-Shuster & None & Exact unconditional test & MTE \\
\midrule
\multicolumn{4}{l}{\textit{Mantel-Haenszel methods}} \\
Cochran-Mantel-Haenszel test & Stratification (categorical) & Stratified $\chi^2$ test & CTE \\
Mantel-Haenszel RD (Sato) & Stratification (categorical) & Wald $z$ test & CPATE \\
Mantel-Haenszel RD (mGR) & Stratification (categorical) & Wald $z$ test & MTE \\
\midrule
\multicolumn{4}{l}{\textit{$g$-computation methods}} \\
$g$-computation (Ge) & Regression model & Wald $z$ test & CPATE \\
$g$-computation (Liu) & Regression model & Wald $z$ test & MTE \\
$g$-computation (Ye) & Regression model & Wald $z$ test & MTE \\
$g$-computation (Steingrimsson) & Regression model & Wald $z$ test & MTE \\
$g$-computation (Zhang) & Regression model & Score test & MTE \\
$g$-computation (Firth) & Penalized regression model & Wald $z$ test & CPATE$^\dag$ \\
\bottomrule
\end{tabularx}
\begin{minipage}{0.95\linewidth}
\footnotesize 
~ \\
Abbreviations:
RD = risk difference;
MTE = marginal treatment effect;
CTE = conditional treatment effect;
CPATE = conditional population-average treatment effect;
mGR = modified Greenland-Robins.
Regression model indicates reliance on a logistic regression working model allowing categorical and/or continuous covariates.
Note, for collapsible measures such as the risk difference, the stratum-wise null tested by the CMH/MH test coincides with the marginal null hypothesis, although the test is derived conditionally. $^\dag$the Firth's regression model does not need to target the CPATE as it is a working model decision rather than an inferential one. In this paper, we use it for the CPATE, however.
\end{minipage}
\end{table}

\subsection{Suissa-Shuster test}

The first approach we consider is the exact unconditional test for comparing two independent binomial proportions proposed by Suissa and Shuster.\cite{Suissa1985} Although it does not accommodate covariate adjustment, it serves as an important benchmark for evaluating the finite-sample properties of covariate-adjusted procedures, particularly with respect to Type~I error control and power. 

The Suissa-Shuster test provides an alternative to \emph{asymptotic} tests (e.g., $z$- or $\chi^2$ tests) and \emph{conditional} tests (e.g., Fisher's exact test\cite{Fisher1935}). Unlike asymptotic tests, it does not rely on large-sample approximations but instead uses the exact finite-sample distribution of the test statistic to maintain the nominal Type~I error rate even in small samples.\cite{Suissa1985} Unlike Fisher's exact test, it does not condition on fixed marginal totals. We include the Suissa-Shuster test because it is constructed directly on the risk-difference scale and therefore aligns the hypothesis test with the estimand of interest. By contrast, other exact unconditional tests, such as the Boschloo test, are based on an ordering induced by Fisher $p$-values and only indirectly connected to inference on the risk difference.\cite{Korn2024}

To outline the test's mathematical underpinnings, recall that we are interested in the difference between two independent binomial probabilities. Let $\pi_1 = \Pr(Y^{(1)} = 1)$ and $\pi_0=\Pr(Y^{(0)} = 1)$ denote the marginal response probabilities under treatment and control, respectively, and $\delta = \pi_1 - \pi_0$ the marginal risk difference. The marginal response probabilities can be reparameterized in terms of $\delta$ and the overall response probability $\theta$: $\pi_1 = \theta + \delta/2$ and $\pi_0 = \theta - \delta/2$. Under the null hypothesis, $\delta = 0$ such that $\pi_1 = \pi_0 = \theta$, and the distribution of observed responses depends only on the nuisance parameter $\theta$.

Let $K_1, K_0$ denote the number of responses and $N_1, N_0$ the corresponding sample sizes. Under the null hypothesis,
\begin{equation*}
    P_{\theta}(K_1, K_0) = \binom{N_1}{K_1} \binom{N_0}{K_0} \theta^{K_1 + K_0}\ (1 - \theta)^{N_1 + N_0 - K_1 - K_0}.
\end{equation*}
The Suissa-Shuster test achieves validity as an \textit{unconditional} test by calculating the $p$~value for an estimated risk difference $\hat \delta_\text{obs}$ as
\begin{equation*}
    p = \sup_{\theta \in [0,1]} \sum_{(k_1^{'},k_0^{'}): \delta(k_1^{'},k_0^{'}) \geq\hat \delta_{\text{obs}}} P_{\theta}(k_1^{'},k_0^{'}),
\end{equation*}
that is, the largest tail probability across all possible $\theta \in [0,1]$, yielding an exact unconditional test. 

Although exact unconditional tests are computationally more demanding than conditional alternatives, this limitation is negligible with modern computing resources. The main limitation is the lack of covariate adjustment, which may reduce efficiency and precludes correction for chance imbalances in prognostic covariates.

\subsection{Mantel-Haenszel methods}

\subsubsection{Cochran-Mantel-Haenszel test}

Cochran-Mantel-Haenszel (CMH) methods\cite{Mantel1959, Agresti2013} are explicitly mentioned in the FDA guidance on covariate adjustment.\cite{FDA2023Covariates} They encompass a family of stratified tests assessing the association between row and column variables across multiple 2$\,\times\,$2 tables (for a comprehensive overview of CMH methods in the context of risk difference, see Qiu et al.\cite{Qiu2025}). 

The test considered here is commonly referred to as \emph{CMH test}, although it is more accurately termed \emph{Mantel-Haenszel (MH) test}.\cite{Qiu2025} Originally proposed as a test of conditional independence across a set of $2 \times 2$ tables, it is used in clinical trials to evaluate whether the null hypothesis of no treatment effect holds uniformly in all of the strata. This null hypothesis can be expressed in terms of a common conditional odds ratio, relative risk, or risk difference. Thus, in the present context, it corresponds to the strict null hypothesis that the conditional risk difference in each stratum is zero, rather than the marginal risk difference across strata. Thus, the CMH/MH test is derived under a conditional framework. 

For collapsible effect measures such as the risk difference, this stratum-wise null implies the marginal null hypothesis of no average treatment effect, even though the test is derived conditionally, whereas for non-collapsible measures the test corresponds strictly to inference on a conditional estimand. This distinction is reflected in the FDA guidance, which endorses CMH methods when interest lies in estimating a conditional treatment effect assumed to be constant across subgroups.\cite{FDA2023Covariates}

For each stratum $s$, let $n_{ijs}$ denote the cell counts in a 2$\,\times\,$2 table denoting the number of observations for each treatment group $i$ (treatment vs. control) and possible outcome $j$ (response vs. no response). Treating the marginal totals as fixed, the test statistic is
\begin{equation*}
    \chi^2_\text{MH} = \dfrac{\left\{ \sum_s n_{11s} - E(n_{11s})\right\}^2 }{\sum_s\text{Var}(n_{11s})},
\end{equation*}
which under the null hypothesis (and conditional on the marginal totals) is asymptotically $\chi^2$ distributed with 1 degree of freedom.

The CMH/MH test is commonly used in clinical trials. It allows for covariate adjustment via stratification. Therefore, adjustment is limited to categorical covariates. Moreover, the validity of the $\chi^2$ approximation requires sufficiently large sample sizes, which is typically assessed using the Mantel-Fleiss criterion.\cite{Mantel1980} 

\subsubsection{Mantel-Haenszel risk difference}

In contrast to the MH test, the Mantel-Haenszel (MH) risk difference estimator is concerned with estimation rather than hypothesis testing and, importantly, targets a different estimand. Although it is often assumed to target a common, conditional treatment effect by averaging stratum-specific risk differences under a homogeneity assumption ($\delta_s = \delta$ for all strata), this strong homogeneity assumption is not required, as recently demonstrated.\cite{Qiu2025} Instead, the MH risk difference estimator can be interpreted as targeting a marginal treatment effect across strata, even in the presence of effect heterogeneity.\cite{Noma2016} Depending on whether covariates are considered fixed or random and on the choice of variance estimator, this average may be viewed as either the CPATE or the MTE; thus, the MH risk difference estimator admits multiple estimand interpretations.

For both estimands, the point estimator is a weighted mean of stratum-specific risk differences:
\begin{equation*}
    \hat \delta = \dfrac{\sum_s w_s \hat \delta_s}{\sum_s w_s},
\end{equation*}
where $w_s$ are stratum-specific weights reflecting sample size and allocation within each stratum.\footnote{Specifically, $w_s = n_{1s}n_{0s}/n_{\cdot s}$, where $n_{1s}=\sum_j n_{1js}$ and $n_{0s}=\sum_j n_{0js}$ denote the sample sizes under treatment and control, respectively, and $n_{\cdot s}=\sum_i\sum_j n_{ijs}$ denotes the total sample size in stratum $s$.}

In contrast, the choice of variance estimator depends on the estimand of interest.
A commonly used variance estimator, proposed by Sato, assumes a common risk difference across strata and ignores between-stratum heterogeneity,\cite{Sato1989} making it appropriate when the CPATE is the estimand of interest. In contrast, Qiu et~al. propose a modified variance estimator that explicitly accounts for heterogeneity in stratum-specific risk differences and is thus appropriate for inference on the MTE.\cite{Qiu2025} 

The situation where a single point estimator may be compatible with multiple estimands while inference depends on the choice of variance estimator is not uncommon and is discussed in detail by Qiu et~al.\cite{Qiu2025} It also arises for the $g$-computation approaches discussed below, and it further speaks to the importance of aligning both estimation and inference.

\subsection{\texorpdfstring{$G$}{G}-Computation}

The $g$-computation (or \textit{standardization}) approach is a general strategy for estimating causal treatment effects by combining an outcome regression model with an explicit averaging step over the covariate distribution.\cite{Robins1986a,Tsiatis2008,VanLancker2024a} In a randomized trial, randomization identifies the marginal mean potential outcomes $E\{Y^{(a)}\}$, which are estimated by predicting outcomes under each treatment level and averaging these predictions over the empirical covariate distribution in the study population. 

In practice, standardization employs a working model for the conditional response mean given treatment assignment $A$ and baseline covariates $\boldsymbol{X}$. For binary endpoints, a common choice is a logistic regression working model,
\begin{equation*}
  \text{logit}\{P(Y=1\mid A,\boldsymbol{X})\}
  = \beta_0 + \beta_A A + \boldsymbol{\beta}_X^{\mathsf{T}}\boldsymbol{X},
\end{equation*}
where $\text{logit}(q)=\log\{q/(1-q)\}$ and $(\beta_0,\beta_A,\boldsymbol{\beta}_X)$ are estimated by maximum likelihood. Under randomization, and for canonical generalized linear models, the resulting estimator of marginal arm-specific means is consistent even under model misspecification.\cite{Moore2009,Ye2023a,VanLancker2024a,Zhang2025}

Given fitted coefficients $\hat\beta_0,\hat\beta_A,\hat{\boldsymbol{\beta}}_X$, the counterfactual (model-predicted) response probability for patient $i$ under treatment level $a\in\{0,1\}$ is
$$
\hat p_i(a) = \text{logit}^{-1}\bigl(\hat\beta_0 + \hat\beta_A a + \hat{\boldsymbol{\beta}}_X^{\mathsf{T}}\boldsymbol{x}_i\bigr).
$$
The standardized estimators of the marginal response probabilities are obtained by averaging these predictions over the observed covariate distribution,
\begin{equation*}
  \hat\pi_a = \frac{1}{N}\sum_{i=1}^N \hat p_i(a), \qquad a\in\{0,1\},
\end{equation*}
and the standardized estimator of the marginal risk difference is
\begin{equation*}
  \hat\delta = \hat\pi_1 - \hat\pi_0.
\end{equation*}
This procedure standardizes to the covariate distribution of the trial population (i.e., the empirical distribution of $\boldsymbol{X}$ in the study sample), yielding a marginal (population-level) estimand with respect to that distribution.\cite{Joffe1995,VanLancker2024a}

While the above procedure describes the standardized point estimator of the marginal treatment effect $\hat\delta$, $g$-computation methods differ primarily in how they estimate $\mathrm{Var}(\hat\delta)$ and construct tests and confidence intervals. A common approach is Wald-type inference based on delta-method variance estimators that treat $\hat\delta$ as a smooth function of estimated regression parameters and account for their estimation uncertainty via sandwich formulas. Early implementations rely on model-based delta-method variances (e.g., Ge et~al.\cite{Ge2011}), whereas more recent work derives model-robust sandwich estimators using stacked estimating equations or influence-function representations, yielding valid inference under randomization even when the outcome working model is misspecified.\cite{Ye2023a,Liu2024,Bannick2025} An alternative class of variance estimators is based on efficient or augmented influence functions for the standardized estimand; these estimators can offer improved efficiency by allowing for machine learning based outcome models but may exhibit non-negligible finite-sample sensitivity, particularly when the working model is poorly behaved.\cite{Zhang2025} As a further alternative, non-parametric bootstrap procedures approximate the sampling distribution of $\hat\delta$ via resampling and are straightforward to implement.\cite{Steingrimsson2017,VanLancker2024a}, however do not necessarily guarantee improved small-sample inference. 
The following subsections provide a concise overview of six selected $g$-computation approaches, outlining their theoretical basis and practical considerations for use in clinical trial settings.

\subsubsection{Ge et al.} 

The widely used approach described by Ge et al.\cite{Ge2011} uses a model-based variance estimator for $\hat \delta$ that treats the observed covariate distribution as fixed. 
Let $\boldsymbol \beta = (\beta_0, \beta_A, \boldsymbol \beta_X)$ denote the parameter vector of the logistic regression working model and $\nabla_{\boldsymbol \beta} \hat \delta$ the gradient of the risk-difference estimator with respect to these parameters, evaluated at $\boldsymbol \beta = \hat{\boldsymbol \beta}$. 
Treating the covariates $\mathbf X$ as fixed, Ge et al.'s variance estimator follows directly from the delta method:
\begin{equation} \label{eq:ge}
    \widehat{\text{Var}}_\text{Ge} (\hat \delta) = \{\nabla_{\boldsymbol \beta} \hat \delta \}^T\ V (\hat{\boldsymbol{\beta}})\ \{\nabla_{\boldsymbol \beta} \hat \delta \},
\end{equation}
where $V (\hat{\boldsymbol{\beta}}) = I(\hat{\boldsymbol{\beta}})^{-1}$ is the inverse observed Fisher information matrix from the fitted model. 
This matrix quantifies the sampling variability of $\hat{\boldsymbol\beta}$ under correct model specification and yields a variance estimate for $\hat\delta$ conditional on the observed covariate distribution.
Consequently, as pointed out by Magirr et al.\cite{Magirr2025}, this approach targets a conditional population-average treatment effect (CPATE).
Consequently, it may underestimate the unconditional variance of $\hat\delta$ because it ignores variability arising from the empirical covariate distribution and potential model misspecification, rendering it less appropriate for inference on the MTE.\cite{Liu2024, Ye2023a}

\subsubsection{Liu and Xi}

Liu and Xi\cite{Liu2024} propose a variance estimator that extends the approach by Ge et al.\cite{Ge2011} to account for model misspecification and target the \emph{unconditional} variance of the $g$-computation estimator. Specifically, they introduce two key modifications: First, they replace the model-based covariance matrix $V(\hat{\boldsymbol\beta})$ with a heteroskedasticity-consistent (sandwich) estimator $V_{\mathrm{HC}}(\hat{\boldsymbol\beta})$, which relaxes the assumption of correct model specification. 
Second, they add a variance component to capture the additional uncertainty arising from the randomness of the empirical covariate distribution. 
Combining these two components yields the following, robust estimator of the unconditional variance:
\begin{equation*}
    \widehat{\text{Var}}_{\mathrm{Liu}}(\hat\delta) = \{\nabla_{\boldsymbol\beta}\hat\delta\}^{T}\, V_{\mathrm{HC}}(\hat{\boldsymbol\beta})\, \{\nabla_{\boldsymbol\beta}\hat\delta\} \;+\; \hat V_{\mathbf{X}}(\hat\delta),
\end{equation*}
where the first term estimates the conditional variance and the second term  estimates the between-sample variance due to the covariate distribution. Consequently, this estimator is appropriate when inference on the MTE is desired.\cite{Liu2024} 
When the conditional risk difference is constant in $\mathbf X$, the additional variance component $\hat V_{\mathbf{X}}(\hat\delta)$ is asymptotically zero. Consequently, in this case, this estimator would agree with Ge et al.'s estimator apart from differences arising from the use of a sandwich covariance estimator. 
Otherwise, ignoring the additional covariate-variability term may lead to systematic underestimation of the unconditional variance and, consequently, anticonservative inference. In their simulation studies, Liu and Xi showed improved Type~I error control relative to conditional variance estimators. For small samples, they recommend the HC3 version of the sandwich estimator, which we also adopt in our simulations.\cite{Liu2024}

\subsubsection{Ye et al.}

Another robust variance estimator for the $g$-computation risk-difference estimator was proposed by Ye et al.\cite{Ye2023a} Their approach is a special case of the general, semi-parametric covariate-adjustment strategy introduced by Bannick et al.\cite{Bannick2025} 
The estimator decomposes the variance into outcome noise, covariate-driven heterogeneity, and their interaction.
Thus, like Liu and Xi's estimator, it targets the unconditional variance, and is therefore appropriate when inference on the MTE is of interest.

\subsubsection{Steingrimsson et al.}

An alternative approach to robust variance estimation for the $g$-computation estimator is the non-parametric bootstrap, as proposed by Steingrimsson et al.\cite{Steingrimsson2017}
This method does not rely on a closed-form variance expression but approximates the sampling distribution of $\hat \delta$ empirically through repeated resampling of the observed trial data. 
For each bootstrap replication $b = 1, \ldots, B$, a sample of size $n$ is drawn  with replacement from the original dataset, the working logistic model is refitted on the bootstrap sample, and the corresponding estimator $\hat \delta^{(b)}$ is computed. The bootstrap variance estimator is then given by the empirical variance of the bootstrap replicates $\{\hat\delta^{(b)}\}_{b=1}^{B}$.

This method yields a model-agnostic variance estimator that accounts for variability due to estimation of the working model, the empirical covariate distribution, and the randomization scheme.
Consequently, it targets the unconditional variance of the treatment effect and is thus appropriate when the MTE is the estimand of interest.
Computationally, the approach is more burdensome than analytic methods, since it requires repeated refitting of the working model. Nevertheless, for a single dataset, it is computationally well feasible on modern hardware, usually completing within a few seconds depending on the number of replicates.

\subsubsection{Zhang et al.}

Zhang et al.\cite{Zhang2025} recently proposed a score-test approach for inference on covariate-adjusted risk differences in the $g$-computation framework.  Motivated by known limitations of Wald-type inference, particularly in small samples or near the boundaries of the parameter space, their method is based on a generalized score test for $M$-estimators\cite{Boos1992} and remains valid under simple and stratified randomization schemes, even when the working model is misspecified.\cite{Zhang2025} 
To test the null hypothesis $H_0\!:\ \delta = \delta_0$, Zhang and colleagues define the test statistic
\begin{equation*}
    \chi^2_\text{Zhang} = \dfrac{(\hat \delta - \delta_0)^2}{\widehat{\text{Var}}(\hat \delta) + \dfrac{(\hat \delta - \delta_0)^2}{n}},
\end{equation*}
where $\widehat{\text{Var}}(\hat \delta)$ denotes any valid variance estimator (as justified in Zhang et al.\cite{Zhang2025}). If an estimator of the unconditional variance is used, as was done in the present simulation study, the approach targets the MTE. 

Compared to the classical Wald statistic, the additional term in the denominator acts as a finite-sample penalty that inflates the variance when $\hat\delta$ deviates from $\delta_0$, thereby stabilizing inference under model misspecification or high-dimensional adjustment.
Under the null hypothesis, the test statistic follows a $\chi^2$ distribution with 1 degree of freedom. 
For estimation, inverting the statistic yields a confidence interval with a simple closed form.

\subsubsection{Firth}

A final approach addresses a practical issue that may arise in small samples, namely, non-convergence of the logistic regression working model due to separation. Separation occurs when (a linear combination of) covariates perfectly predict the outcome, in which case maximum likelihood estimates do not exist in the usual sense and numerical optimization fails. 

To address this issue, we consider a penalized-likelihood approach that combines the bias-reduction procedure of Firth\cite{Firth1993} with the FLIC modification proposed by Puhr et al.\cite{Puhr2017} in order for the predictions to match the marginal means without the penalization. Firth's method augments the logistic regression likelihood with a Jeffreys prior penalty to reduce small-sample bias. 
While that approach stabilizes parameter estimation, it induces shrinkage of predicted probabilities toward $0.50$. The FLIC (``\textit{Firth Logistic regression with Intercept Correction}'') modification counteracts this by adjusting the estimated intercept so that the average predicted probability matches the observed event rate, while keeping the remaining coefficient estimates fixed. This post-hoc adjustment improves calibration without sacrificing the bias-reduction and stability properties of Firth's method.\cite{Puhr2017} 

In the $g$-computation framework, the penalized model with FLIC correction can be used as a drop-in replacement for the standard working model. 
The point estimator of the marginal risk difference is then obtained via the usual standardization procedure from the fitted linear predictor under each treatment arm. For inference on the estimated risk difference, we pair the penalized working model with the model-based conditional variance estimator by Ge et al., which yields inference on the CPATE, although any valid variance estimator could in principle be used.\cite{Ge2011}

\section{Simulation design}
\label{sec:simstudy}

\subsection{Settings}

The simulation design mirrors the hypothetical clinical trial introduced in Section~\ref{sec:hypotheticaltrial}. We considered a binary endpoint $Y$ and three independent predictor variables: the randomized treatment indicator $A$ and two binary baseline covariates $X_1$ and $X_2$. Treatment allocation followed simple randomization with a 1:1 allocation ratio. The covariates $X_1$ and $X_2$ were generated from independent Bernoulli distributions with success probability $0.50$.
The binary response $Y$ was generated via a multivariable logistic regression model including treatment and both covariates:
$$ 
\text{logit}\{P(Y = 1 \mid A, X_1, X_2)\} =
\beta_0 + \beta_A A + \beta_1 X_1 + \beta_2 X_2.
$$

This data-generating model matches the working model used for the $g$-computation approaches and aligns with the covariate structure used in the MH test and MH risk-difference estimator. Hence, all methods are evaluated under correctly specified conditions. Since model misspecification has already been studied extensively elsewhere (e.g., by Liu and Xi\cite{Liu2024}), we did not incorporate additional misspecification scenarios.

To control both the marginal response probability under control and the marginal risk difference, we determined $\beta_0$ and $\beta_A$ using the iterative bisection algorithm proposed by Austin.\cite{Austin2023} The marginal response probability under control was fixed at $0.20$, while the marginal risk difference $\delta$ was varied across three scenarios: $\delta = 0$ (no treatment effect), $\delta = 0.15$ (moderate effect), and $\delta = 0.30$ (large effect). The prognostic strength of the covariates was varied by setting: $\beta_1 = \beta_2 \in \{\log(1),\ \log(1.5),\ \log(3)\}$. The resulting conditional and marginal response probabilities across all parameter settings are shown in Table \ref{tab:simRespProbs}.

To explore performance under small-sample conditions, the total sample size was varied across five levels, with $N \in \{30, 60, 90, 120, 150\}$. In total, the simulation comprised $3$ (treatment effects) $\times\ 3$ (covariate prognostic strengths) $\times\ 5$ (sample sizes) $= 45$ scenarios. For each scenario, $50{,}000$ trial datasets were generated, and all ten methods were applied to each dataset.

Within our simulation framework, the data-generating mechanism was constructed such that the simulated true marginal risk difference corresponds to the MTE. Consequently, the performance of all statistical methods is evaluated with respect to this estimand, even though some of them are designed to target alternative estimands, as discussed above. Given that the MTE is often of primary interest in clinical trials, this choice is informative and practically relevant. At the same time, differences in the targeted estimands must be considered when interpreting results, as observed discrepancies in performance may, at least partly, reflect estimand mismatch rather than inherent shortcomings of the methods.

Finally, note that our data-generating mechanism was intentionally constructed not to include longitudinal measurements and missing outcomes. As a result, the simulation does not allow evaluation of the methods' performance under different missing-data mechanisms or in conjunction with missing-data handling procedures, such as multiple imputation. While these considerations may be relevant for method selection in longitudinal clinical trials, they introduce additional complexities that are separate from the primary objective of this work and, thus, beyond the scope of the current simulation design.

\subsection{Performance measures}

To compare the statistical methods across the simulation settings described above, we evaluated a set of operating characteristics. Because our primary interest lies in statistical inference in the form of hypothesis testing -- assessing whether a treatment effect exists -- the central metric is the rejection rate. 
Under the null hypothesis ($\delta = 0$), this corresponds to the Type I error rate, whereas under alternative scenarios with $\delta \neq 0$ it reflects statistical power.

In addition we examined the quality of point and interval estimation for all methods, excluding the MH test, which does not afford an estimate of the risk difference. Estimation bias was quantified as the absolute deviation between the average estimated and the true risk difference, and precision was assessed using the root mean squared error (RMSE). We further evaluated the empirical coverage probabilities of nominal 95\% confidence intervals (CIs). Together, these metrics characterize estimator performance beyond hypothesis testing.

Finally, we investigated computational challenges particularly in the $g$-computation approaches. Logistic regression working models may suffer from separation and non-convergence in small samples, particularly when covariates are strongly prognostic or the number of covariates is large relative to the sample size. Such issues may impair feasibility, induce missingness, and potentially distort conclusions about the inferential properties of the investigated methods, yet are rarely reported in simulation studies.\cite{Pawel2026}
We therefore quantified the frequency of computational issues across all simulation scenarios and examined their downstream impact on the methods' inferential characteristics. Thereby, we seek to ensure that our performance assessment reflects not only theoretical behavior under ideal conditions but also practical limitations encountered in small-sample trial settings.

\section{Results}
\label{sec:results} 

\subsection{Computational issues}

Before evaluating operating characteristics and estimation performance, we examined potential computational issues associated with the statistical procedures under consideration, with particular focus on convergence and separation in the logistic regression working models used in the $g$-computation approaches. 
In the presence of complete or quasi-complete separation, the maximum likelihood estimate of some model parameters does not exist. In such cases, numerical optimization cannot converge as the algorithm attempts to approach infinite parameter values. However, depending on the convergence criterion, the algorithm may still formally declare convergence. 
We therefore distinguish between explicit non-convergence of the numerical algorithm, where no parameter estimates or predicted values are obtained, and separation with formal convergence, where estimates are returned despite the absence of a finite maximum likelihood solution. 

Non-convergence was defined using the default criterion of the \texttt{glm} function in R, where the iterative reweighted least squares algorithm is deemed to have converged if the relative change in deviance falls below $10^{-8}$ within a maximum of 25 iterations; otherwise, non-convergence is declared. Separation with formal convergence was detected using the \texttt{detectseparation} package\cite{detectseparation}, which provides post-estimation diagnostics for identifying separation and infinite maximum likelihood estimates in generalized linear models.\cite{Konis2007} 

\begin{figure}[hb]
    \centering
    \includegraphics[width=0.7\textwidth]{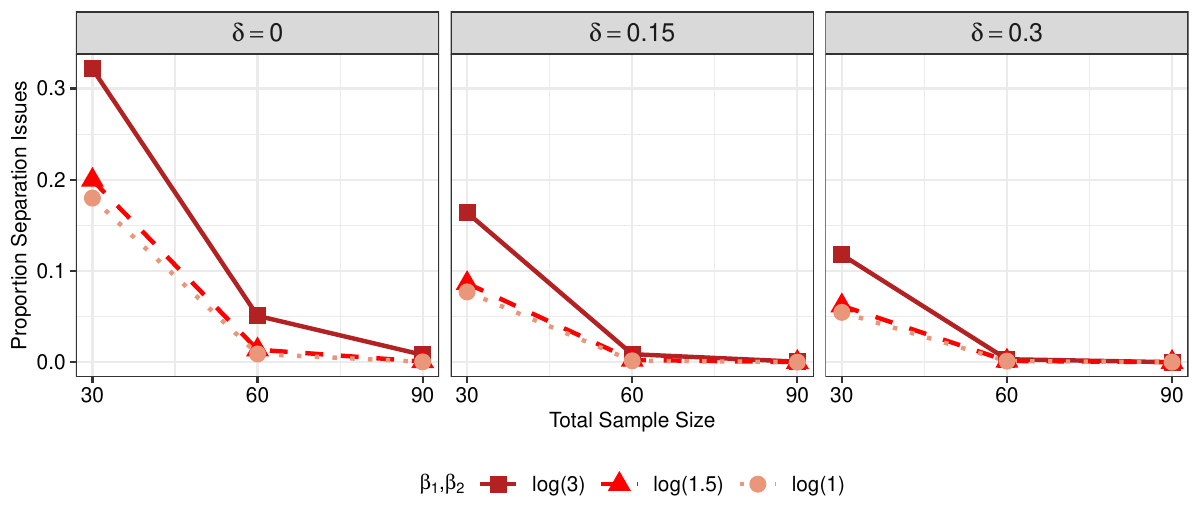}
    \caption{Proportion of simulation runs with separation issues as indicated by the \texttt{detectseparation} package. Separation did not occur in sample sizes larger than 90.}
    \label{fig:separation}
\end{figure}

Non-convergence was rare across all scenarios. Only $0.0036\%$ of the simulation runs failed to converge, all at the smallest sample size ($N = 30$). 
We excluded all simulation runs that ended in formal non-convergence before analysis. Given their negligible frequency, this is unlikely to materially affect the reported operating characteristics. In contrast, separation with formal convergence occurred much more frequently (Figure \ref{fig:separation}). For small samples ($N = 30, 60$) and strongly prognostic covariates, separation was observed in up to over 30\% of simulations. This incidence decreased with increasing sample size and larger treatment effects. Nonetheless, in realistic trial settings with small samples and strongly prognostic covariates, separation remains a non-negligible concern. 

When separation is driven by covariates, one possible mitigation strategy is to remove problematic predictors, as suggested for example by Liu and Xi.\cite{Liu2024} However, our results indicate that separation attributable to covariates did not materially affect performance of the $g$-computation approaches. In contrast, separation driven by the treatment indicator reflects an extreme observed treatment effect and typically affects all methods, often resulting in inflated Type~I error rates. In such cases, excluding covariates does not resolve the issue. For these reasons, we retained all formally converged simulation runs regardless of the presence of separation.

Computational anomalies were not confined to regression approaches. The MH test occasionally failed when only a single outcome category was observed across strata, causing the variance estimator to collapse and preventing computation of the test statistic. This occurred in approximately $0.16\%$ of runs, exclusively at $N = 30$, and these runs were excluded. Overall, computational issues were rare for most methods. Separation in logistic regression emerged as the primary computational challenge in small-sample settings and warrants careful consideration when interpreting inferential results.

\subsection{Type I error rate}  

\begin{figure}[ht]
    \centering
    \includegraphics[width=\textwidth]{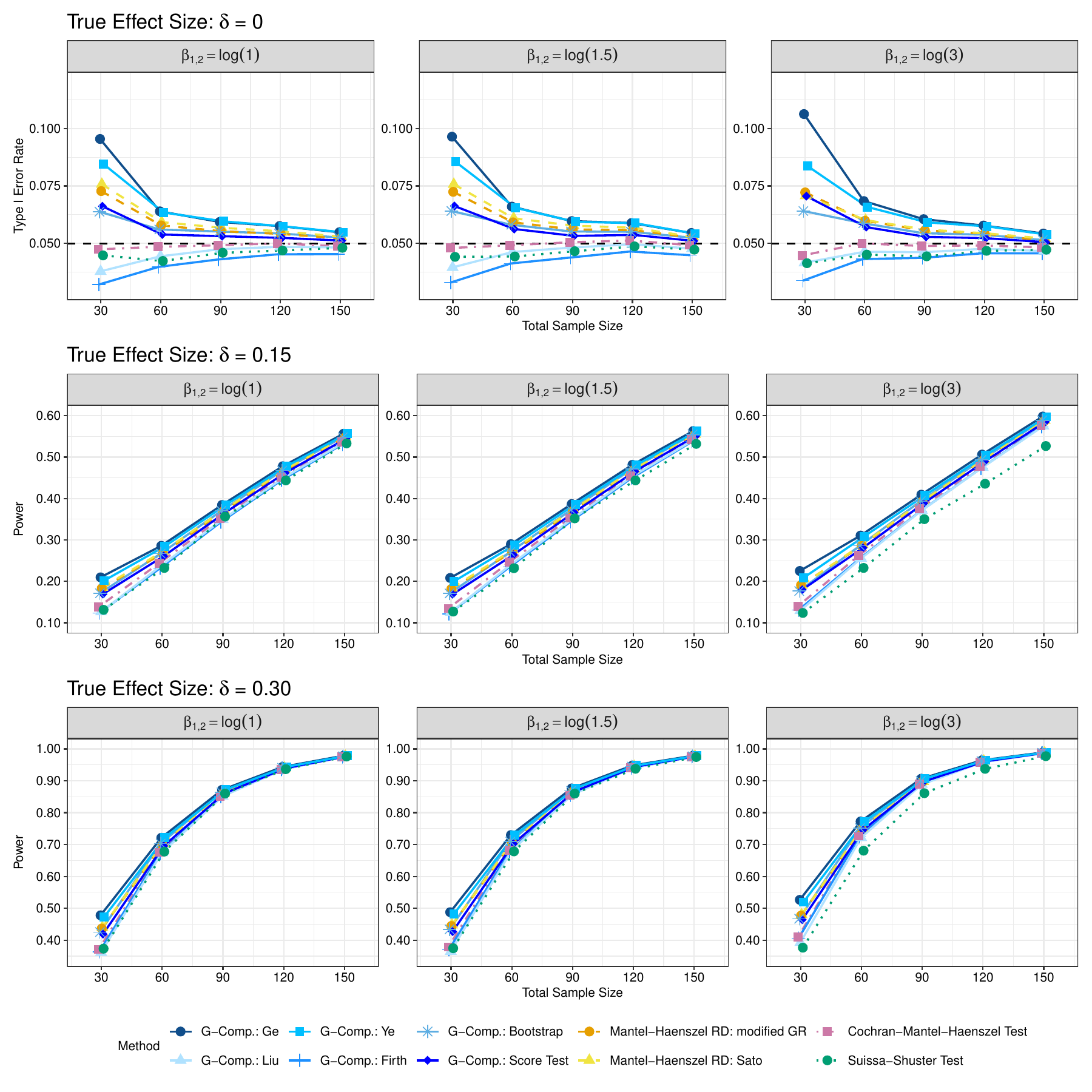}
    \caption{Type I error rates (top row) and statistical power (middle and bottom row) for the investigated statistical methods across simulation scenarios. Note that $y$-axes differ across rows.}
    \label{fig:ocs}
\end{figure}

Across all scenarios with a true risk difference of zero, the observed Type~I error rates exhibit a consistent pattern driven primarily by sample size and, to a lesser extent, by the prognostic strength of the included covariates (Figure~\ref{fig:ocs}, top row). 

At the smallest sample size ($N = 30$), several $g$-computation approaches show pronounced Type~I error inflation, particularly those relying on the (semi-)parametric variance estimators proposed by Ge et al.\ and Ye et al. Importantly, this inflation reflects variance underestimation rather than a failure of $g$-computation per se. The $g$-computation approaches based on the non-parametric bootstrap and the score test exhibit slightly smaller, but still clearly inflated, Type~I error rates. In contrast, the approach by Liu and Xi and the Firth-based $g$-computation method behave conservatively, with error rates well below $0.05$ in very small samples. As sample size increases, error rates for all $g$-computation methods steadily converge toward the nominal level, with most approaches falling within an acceptable range around $0.05$ by $N \geq 120$.

For the MH risk-difference estimator, the variance estimators proposed by Sato et al.\ and Qiu et al.\ display highly similar operating characteristics across all scenarios. Both exhibit clear Type~I error inflation in small samples, slightly higher than those of the bootstrap-based $g$-computation approach. This inflation decreases with increasing sample size and approaches the nominal level by $N \geq 120$. As for the other methods, convergence toward nominal Type~I error control is faster when covariates are more strongly prognostic.

The MH test maintains Type~I error control close to the nominal level across all sample sizes and covariate settings, with only minor fluctuations around $0.05$. The Suissa-Shuster exact unconditional test is consistently the most conservative procedure, with Type~I error rates below $0.05$ across all scenarios, although these gradually approach the nominal level as sample size increases.

Taken together, these results suggest that (i) many $g$-computation approaches are liberal in very small samples irrespective of covariate prognostic strength, (ii) robust or penalized variants, such as those proposed by Liu and Xi or based on Firth's correction, mitigate Type~I error inflation at the cost of conservatism, and (iii) the MH and Suissa-Shuster tests remain robust, with the latter being characteristically conservative.

\subsection{Power}

Simulation results for statistical power are shown in the middle and bottom rows of Figure~\ref{fig:ocs}. Power was evaluated under two treatment-effect scenarios: a small effect ($\delta = 0.15$) and a large effect ($\delta = 0.30$). For the smaller effect size, power was uniformly low across all methods, increasing from approximately $0.10$ at $N = 30$ to about $0.60$ at $N = 150$. Although such operating characteristics are clearly suboptimal and indicate substantially underpowered settings, they still allow for a comparison of \emph{relative} performance of competing approaches. For the larger effect size, power covered a more practically relevant range, increasing from moderate levels at $N=60$ to nearly $0.99$ for $N=150$.

Among the $g$-computation approaches, the (semi-)parametric variance estimators proposed by Ge et al.\ and Ye et al.\ consistently achieved the highest power. These gains, however, came at the cost of inflated Type~I error in small samples. At the other extreme, the more conservative procedures based on a Firth-penalized logistic regression working model and the variance estimator of Liu and Xi exhibited the lowest power in small samples. Bootstrap- and score-test based approaches showed intermediate performance, reflecting their moderate Type~I error inflation. As sample size, treatment effect, and covariate prognostic strength increased, power rose rapidly for all methods and differences between approaches diminished.

For the MH risk-difference estimator, the variance estimators proposed by Sato et al.\ and Qiu et al.\ again showed nearly identical performance. Their power was broadly comparable to $g$-computation with bootstrap or score-test inference. The MH test exhibited power similar to that of the Liu-Xi and the Firth-based $g$-computation approaches. In the settings considered here, and under a requirement for strict Type~I error control, these results suggest that $g$-computation does not provide a substantial power advantage over the MH test. 
However, two qualifications are important. First, the MH test targets a different estimand than $g$-computation, namely a common conditional treatment effect. Second, the present simulations considered only categorical covariates. As noted by Ge et al., efficiency gains of $g$-computation over CMH-type methods become more pronounced when continuous covariates are available, as these would need to be discretized to be accommodated by CMH approaches.\cite{Ge2011}

Finally, the Suissa-Shuster exact unconditional test consistently yielded the lowest power across all scenarios. The gap relative to covariate-adjusted methods reflects the absence of covariate adjustment, and it increases as covariates become more prognostic.

Overall, the results indicate that (i) for small samples and moderate effects, power is low for all methods. The most powerful procedures achieve gains at the cost of substantial Type~I error inflation, whereas penalized or robust variants are more conservative and therefore less powerful. (ii) As the treatment effect, covariate prognostic strength, and sample size increase, power improves rapidly and differences between methods decrease. In terms of power, MH risk-difference estimators perform similarly to $g$-computation using bootstrap or score-test inference, while the MH test aligns with more conservative $g$-computation variants. Finally, (iii) method selection should consider both the targeted estimand and the covariate structure. When only categorical covariates are available and strict Type~I error control is essential, the MH test provides a robust baseline. In contrast, the relative power advantages of $g$-computation can be expected to be more pronounced when continuous covariates are available.

\subsection{Estimation bias}

Beyond hypothesis testing, precise estimation of the treatment effect is often of key interest in clinical trials, particularly in non-confirmatory early-phase studies. To evaluate estimation performance, we assessed bias, RMSE, and coverage of 95\% CIs for all methods that provide a risk-difference estimate. Let $\hat\delta_r$ denote the estimate in replication $r = 1,\ldots,R$, and $\delta$ the true value. Bias and RSME are then defined as
$$
  \text{Bias} = \left\{ \frac{1}{R} \sum_{r=1}^R \hat\delta_r \right\} - \delta \quad \text{and} \quad 
  \text{RMSE} = \sqrt{\dfrac{1}{R} \sum_{r=1}^R (\hat{\delta}_r - \delta)^2 },
$$
respectively. Results for all methods under all scenarios are displayed in Figures \ref{fig:bias} and \ref{fig:rmse}. 

Under the null hypothesis ($\delta = 0$), all estimators are approximately unbiased.
When $\delta > 0$, this remains the case for all methods but the Firth-based $g$-computation estimator. This behavior is expected as the Firth correction induces shrinkage of predicted probabilities toward $.50$. At the same time, variance is reduced, leading to lower RMSE compared to other methods. Both the bias and the RMSE advantage decrease as sample size increases.

\begin{figure}[ht]
    \centering
    \includegraphics[width=\textwidth]{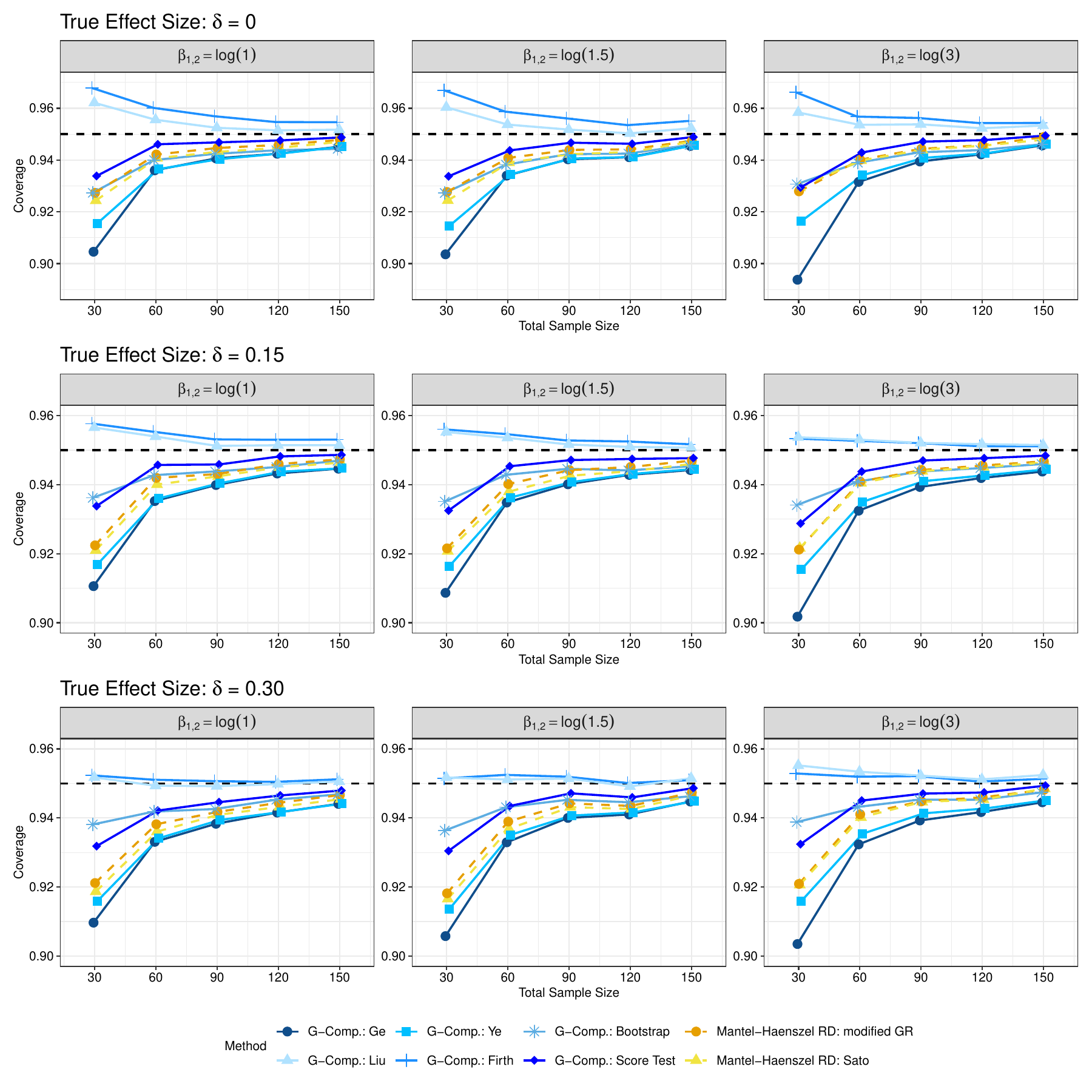}
    \caption{Coverage of two-sided 95\% confidence intervals for all considered statistical methods.}
    \label{fig:coverage}
\end{figure}

To evaluate interval estimation, we examined empirical coverage of two-sided 95\% CIs (Figure \ref{fig:coverage}). Except for the score-test-based approach, where confidence limits were obtained by inverting the $\chi^2$ test statistic, all methods use Wald-type intervals.

Under the null hypothesis, coverage rates simply reflect observed Type~I error rates. Under non-zero treatment effects, the overall pattern remains largely consistent: The Ge-, Ye-\, MH-, and score-test-based estimators continue to show undercoverage, whereas the bootstrap-based intervals improve modestly, achieving coverage closer to the nominal level. Finally, the Liu- and Firth-based approaches demonstrate excellent coverage performance across effect sizes, with empirical coverage rates very close to the nominal 95\%. Thus, despite their conservative behavior and reduced power, these methods provide reliable interval estimates when a true treatment effect is present even in small to moderate samples.

\section{Discussion}
\label{sec:discussion}

\subsection{Main insights}

Invigorated by recent regulatory guidance, there has been growing interest in covariate-adjusted assessments of marginal risk differences in clinical trials with binary endpoints.\cite{FDA2023Covariates} This development, in turn, has fostered a strong focus in the methodological literature on $g$-computation, a flexible framework for estimating causal effects with non-linear endpoints and both continuous and discrete covariates. Much of the existing literature, however, centers on asymptotic theory, leaving the finite-sample performance of these methods, and particularly their behavior in small samples, less well understood. This gap has been explicitly noted in recent work, which highlights the need for further research (e.g., Bannick et al.\cite{Bannick2026}). 

To address this need, we conducted a comprehensive simulation study comparing ten statistical procedures for the assessment of risk differences within the context of small-sample clinical trials. 
Our study yields several insights: First, non-convergence of logistic regression models was rare, even in very small sample sizes. In contrast, separation with formal convergence occurred with appreciable frequency in extremely small samples. Importantly, though, its negative impact on downstream performance was limited. Nevertheless, practitioners should remain aware of potential computational anomalies and consider robust alternatives such as a Firth-penalized logistic regression\cite{Firth1993,Puhr2017} or assumption-lean procedures like the Suissa-Shuster test\cite{Suissa1985} when necessary.

Second, our results revealed considerable heterogeneity in inferential performance across methods. Several approaches exhibited marked inflation of Type I error rates in small samples, most prominently $g$-computation with the variance estimators proposed by Ge et al.\ and Ye et al. It is important to keep in mind, however, that our simulation framework evaluated inferential properties with respect to the MTE, which is not the targeted estimand for all of the methods considered. For example, $g$-computation with the variance estimator of Ge et al. targets the CPATE. Thus, the observed error inflation for this approach may at least partly reflect estimand mismatch (see also Magirr et al.\cite{Magirr2025}).
The MH risk difference estimator and score-test- and bootstrap-based approaches also showed elevated error rates, though all methods converged quickly toward nominal levels as sample sizes increased. 
In contrast, both the Suissa-Shuster and the MH test demonstrated excellent to conservative Type I error control across all scenarios, though at the cost of smaller power. 

Finally, robust $g$-computation approaches using the variance estimator of Liu and Xi or a Firth-penalized logistic regression model exhibited conservative Type I error behavior, particularly in very small samples, and consequently reduced power, but excellent coverage. As expected, the Firth-based approach introduced a characteristic downward bias when the true risk difference was non-zero. Nevertheless, both methods delivered the most reliable interval estimates under small-sample conditions.

Our findings complement earlier work demonstrating the efficiency benefits of standardization-based estimators under randomization.\cite{Colantuoni2015} While these results establish consistency and asymptotic efficiency of marginal estimators, our simulations show that in small samples, inferential validity hinges critically on aligning the variance estimator with the estimand. Our findings are consistent with recent comparative work showing that, in randomized trials, covariate adjustment can improve efficiency but inferential performance, particularly variance and interval estimation, depends critically on modeling assumptions.\cite{Tackney2023}

\subsection{Practical recommendations}

Our findings reveal an unsettling, yet not unexpected insight: No single statistical method dominates across all relevant performance dimensions. Instead, the choice of a method necessarily entails balancing competing objectives such as computational stability, Type~I error control, statistical power, and the accuracy and precision of treatment-effect estimation. These characteristics must be weighed carefully in light of the specific trial objectives, particularly the estimand of interest and the role of prognostic covariates.

When strict Type I error control for inference on the MTE is essential, several approaches emerge as viable options. The Suissa-Shuster test maintains robust error control across all scenarios, albeit at the cost of not leveraging prognostic covariates and, consequently, reduced power. The appeal of such unconditional, design-based procedures lies in their transparency and minimal reliance on modeling assumptions, a perspective emphasized in critiques of regression adjustment in randomized experiments.\cite{Freedman2008} Alternatively, to leverage prognostic information from categorical covariates, the CMH test offers similarly reliable error control. Its applicability is limited, however, if sparse data result in too small cell counts or in the presence of continuous covariates.\cite{Mantel1980} For these cases, robust $g$-computation combined with the Liu and Xi variance estimator or a Firth logistic regression working model offers a principled alternative. These methods provide conservative error control in small samples at the cost of reduced statistical power, although the advantage of covariate adjustment becomes evident with moderate sample sizes ($N \geq 90$).

When balance between Type I error control and power is desired, $g$-computation with either score-test-based\cite{Zhang2025} or bootstrap-based\cite{Steingrimsson2017} inference provides an appealing compromise. Both approaches exhibit less Type~I error inflation than the MH risk-difference estimator while achieving comparable power. 
When continuous prognostic covariates are present, we anticipate the efficiency gain of these $g$-computation methods to be even more pronounced.

When the primary focus is on estimation rather than hypothesis testing, considerations shift toward bias and variance. Most evaluated methods yield unbiased estimators, except for $g$-computation with a Firth logistic regression working model, which introduces shrinkage-induced bias in point estimation but benefits from reduced variance and improved RMSE, as well as excellent confidence-interval coverage. Liu and Xi's approach similarly provides very strong coverage performance without introducing bias. 

\subsection{Consistency between estimation, testing, and confidence intervals}

From a regulatory and scientific reporting perspective, it is  desirable that a single analytical framework yields a coherent trio consisting of (i) a point estimator, (ii) a hypothesis test, and (iii) a confidence interval that are all aligned with the same estimand and based on compatible inferential logic. Inconsistencies between these components, such as testing one estimand while estimating another, or reporting confidence intervals that are not the inversion of the primary test, can complicate interpretation, undermine transparency, and raise concerns in regulatory review. Unadjusted and stratified design-based procedures illustrate these limitations clearly. In contrast, $g$-computation provides a unifying framework in which estimation, testing, and interval estimation are all derived from the same underlying standardized estimand. The standardized risk-difference estimator $\hat\delta = \hat\pi_1 - \hat\pi_0$ is well defined, interpretable on the marginal scale, and applicable to arbitrary covariate structures. The remaining challenge is how to construct valid inference around this estimator in finite samples. Among the explored approaches, two stood out as providing internally consistent inference within the $g$-computation framework: robust variance-based Wald inference targeting the MTE, as in Liu and Xi,\cite{Liu2024}, and score-test-based inference, as proposed by Zhang et al.\cite{Zhang2025} The principal value of $g$-computation in small-sample trials is not adjustment per se, but coherence. When paired with an inferential procedure that respects the estimand and accounts for finite-sample uncertainty, $g$-computation offers a principled and credible route to reporting risk differences that integrates estimation, testing, and uncertainty quantification within a single framework.

\subsection{Limitations and future directions}

There are limitations of our study that merit consideration and point toward avenues for future research. First, our simulations considered only categorical baseline covariates. This choice ensured a controlled comparison across methods, where performance differences cannot be attributed to differences in the set of considered covariates, but it necessarily restricts the generalizability of our findings. In practice, continuous prognostic covariates are common, and their inclusion may alter the relative performance of the approaches considered. In particular, efficiency gains of $g$-computation relative to CMH-type procedures will be more pronounced when continuous covariates can be fully exploited without discretization (see Ge et al.\cite{Ge2011}). Future work may also consider scenarios with more extreme outcome probabilities, such as marginal risks close to 0 or 1, where asymptotic approximations are known to break down and differences between methods may be amplified.

Second, we restricted the simulated scenarios to correctly specified models in the sense that all covariate-adjusted procedures included the covariates used for data generation. This design choice ensured a level playing field across approaches. Nevertheless, model misspecification is an omnipresent concern in applied analyses.\cite{Tackney2023} The relative robustness of the methods -- particularly $g$-computation variants -- to omitted covariates, incorrect link functions, or unmodeled interactions thus remains to be addressed.

A closely related aspect concerns the choice of estimand in the simulation design. The data-generating mechanism was specified such that the MTE is controlled, reflecting that the considered methods are commonly applied to target this estimand. At the same time, some methods are designed to target alternative estimands, such as the CPATE or a common CTE(x), so the simulation deliberately allows for estimand mismatch. Consequently, part of the observed performance differences may reflect this mismatch rather than method-specific deficiencies. Future studies could explicitly control alternative treatment effects to disentangle the impact of estimand mismatch from method-specific performance issues.

Third, our simulations were based on simple randomization, which is arguably not the most prevalent choice in clinical trials. Stratified or covariate-adaptive randomization schemes are increasingly used to balance treatment assignment across key covariates. Variance estimators that appropriately account for stratified randomization are available, e.g., Wang et al.\cite{Wang2023}, Bannick et al.\cite{Bannick2025}, or Zhang et al.\cite{Zhang2025}, and may further improve efficiency. Future work should evaluate the impact of alternative randomization schemes and corresponding inference strategies. In addition, extensions to unequal allocation ratios (i.e., designs other than 1:1 randomization) would be of practical relevance, as these may affect both efficiency and small-sample behavior.

Fourth, finite-sample corrections for Wald-type inference based on sandwich variance estimators, such as the bias-adjusted variance and $F$-reference distributions proposed by Fay and Graubard,\cite{Fay2001} could in principle be applied to $g$-computation estimators. However, they require nontrivial choices regarding the effective number of independent contributions and parameter-specific degrees of freedom, and are not routinely used. Similarly, Tsiatis et al.\ note that sandwich-based variance estimators for covariate-adjusted estimators can be downward biased in small samples and propose multiplicative variance correction factors based on sample size and model complexity.\cite{Tsiatis2008} A systematic evaluation of such corrections for marginal risk differences remains an open area for future research.

Beyond these corrections, alternative strategies to improve inference on the risk-difference scale warrant investigation. Variance-stabilizing transformations (e.g., arctanh transformations) may improve the behavior of Wald-type procedures near the boundary of the parameter space. More generally, the scale dependence of Wald tests suggests that inference on alternative measures, such as odds ratios or risk ratios, may yield improved finite-sample properties. Likewise, higher-order methods such as double-bootstrap procedures may provide more accurate coverage and error control in small samples, but come at increased computational cost.

Finally, to maintain a focused comparison, several relevant approaches were not included in the present study. In particular, inverse probability of treatment weighting (IPTW), augmented inverse probability weighting (AIPW), and targeted maximum likelihood estimation (TMLE) were not considered.\cite{Williamson2014,Moore2009,Rosenblum2010} Previous work indicates that robust standard errors for IPTW were anti-conservative in small samples,\cite{Morris2022,Tackney2023} while the small-sample performance of AIPW and TMLE warrants further investigation.
Related approaches, such as overlap weighting, doubly robust estimating equations, and higher-order influence-function estimators, were also omitted as their implementation and inferential targets would require separate, dedicated investigation. Future work should also consider the approach proposed by Klingenberg,\cite{Klingenberg2014}  as well as recent developments for $g$-computation by Lee and colleagues,\cite{Lee2026} who propose an influence-function-based leave-one-out cross-validated variance estimator that has shown favorable properties in settings with small samples or rare outcome events. Extending the present simulation framework to include such approaches would provide an even more comprehensive assessment of covariate-adjusted inference for binary endpoints.

\subsection{Conclusion}
This study provides a large simulation-based evaluation of methods for covariate-adjusted inference on risk differences in small-sample randomized clinical trials. Rather than identifying a universally superior procedure, our results show that each has a distinct profile of advantages and disadvantages with respect to computational stability, Type I error control, statistical power, and estimation accuracy. Simple, highly robust procedures such as unconditional exact tests ensure strict error control but forego the efficiency gains available through covariate adjustment, whereas $g$-computation methods can capitalize on prognostic information yet require careful selection of variance estimators and working models to perform reliably in small samples. Our findings underscore the importance of selecting statistical methods that are explicitly aligned with the estimand of interest, the inferential objective, and practical trial constraints. When this alignment is achieved, covariate-adjusted analyses can deliver coherent point estimates, tests, and confidence intervals that are consistent with modern estimand-focused regulatory guidance. Overall, these results reinforce that method selection should be guided by the estimand and inferential framework, rather than by covariate adjustment alone. By clarifying the trade-offs and providing empirically grounded guidance, this work supports principled and transparent analysis choices for small-sample clinical trials.

\subsection*{Conflicts of Interest}

Martin Schnuerch and Christian Stock are employees of Boehringer Ingelheim Pharma GmbH \& Co. KG, Germany. Alex Ocampo is an employee of F. Hoffmann-La Roche Ltd, Basel, Switzerland. Klaus Kähler Holst is an employee of Novo Nordisk, S\o{}borg, Denmark.

\subsection*{Data availibility statement}

The simulated raw data underlying all results and figures reported in this manuscript are archived on Zenodo and publicy available from \url{https://doi.org/10.5281/zenodo.21371937}. The data and all results are fully reproducible. Simulations were conducted in the \textsf{R} statistical computing environment (version~4.5.2) using a high-performance computing infrastructure managed by the Slurm Workload Manager. The \textsf{R} scripts used to generate the simulated data, perform the analyses, and produce the figures and tables, together with documentation and replication instructions, are made available via a GitHub repository at \url{https://github.com/Boehringer-Ingelheim/binary-endpoints}. Further details on the employed specialized packages and functions used to implement the investigated methods are given in Appendix~\ref{app:software}.

\bibliography{refs, r-packages}{}
\bibliographystyle{ieeetr}

\appendix

\section{Software implementation}
\label{app:software}

This appendix summarizes practical aspects of software implementation for the statistical methods described and discussed in the main text. The focus is on commonly used and well-maintained \textsf{R} packages that were also employed in our simulation study, with brief reference to alternative implementations. The listed packages and functions are not intended to provide an exhaustive overview of available software.

\subsection{Exact unconditional testing}

The Suissa-Shuster exact unconditional test for comparing two independent binomial proportions is implemented in \textsf{R} through the \texttt{Exact} package.\cite{Exact} The function \texttt{exact.test()} from that package provides direct support for unconditional tests on the risk difference scale.

\subsection{Mantel-Haenszel methods}

Mantel-Haenszel tests and Mantel-Haenszel risk difference estimators are available in standard statistical software. In \textsf{R}, the CMH test can be carried out using the \texttt{mantelhaen.test()} function from the \texttt{stats} package. Estimation of the Mantel-Haenszel risk difference and associated variance estimators can be implemented using custom code, the \textsf{R} package \texttt{RobinCar}\cite{RobinCar}, or via routines in \textsf{SAS} (\texttt{PROC FREQ}), where both the classical and modified Greenland-Robins variance estimators are available.

\subsection{\texorpdfstring{$G$}{G}-computation}

\subsubsection{Model-based variance estimation}

For practical applications in which a conditional estimand such as the CPATE is appropriate, $g$-computation combined with the model-based variance estimator of Ge et al.\ can be implemented straightforwardly using standard regression output.\cite{Ge2011} The authors provide working examples in both \textsf{SAS} and \textsf{R} in the appendix of their publication. In \textsf{R}, this approach can be implemented without custom code for variance estimation using the \texttt{margins} package, which computes standardized (marginal) predictions from generalized linear models together with delta method standard errors based on the model-based covariance matrix of the fitted regression coefficients.\cite{margins} This implementation treats the empirical covariate distribution as fixed and yields inference conditional on the observed covariates. The approach by Ge et al.\ is also implemented in the \textsf{R} package \texttt{beeca}.\cite{beeca}

\subsubsection{Robust variance estimation}

Robust variance estimators for $g$-computation targeting the marginal treatment effect are available through several \textsf{R} packages. The variance estimator proposed by Liu and Xi can be implemented using a combination of the \texttt{margins} and \texttt{sandwich} packages, replacing the model-based covariance matrix with a heteroskedasticity-consistent estimator and adding the covariate-distribution variability term.\cite{margins, sandwich,Liu2024} The authors provide \textsf{R} code illustrating the implementation.

The variance estimator of Ye et al.\ is implemented in the \textsf{R} packages \texttt{RobinCar}\cite{RobinCar}, \texttt{RobinCar2}\cite{RobinCar2}, and \texttt{beeca}.\cite{beeca}

\subsubsection{Bootstrap-based}

Non-parametric bootstrap inference for $g$-computation can be implemented using base \textsf{R} functionality in conjunction with standard generalized linear model fitting routines. Steingrimsson et al.\ provide reference implementations in both \textsf{R} and \textsf{SAS} in their supplementary materials.\cite{Steingrimsson2017} The bootstrap procedure requires repeated refitting of the working outcome model and recomputation of standardized risk differences across bootstrap samples.

\subsubsection{Score-test-based inference}

The score-test-based procedure proposed by Zhang et al.\ can be implemented using fitted values from a standard logistic regression working model and any valid variance estimator for the standardized risk difference.\cite{Zhang2025} The authors provide reference \textsf{R} code in their supplementary materials. The implementation does not require numerical inversion of likelihoods and relies only on closed-form expressions for the test statistic and corresponding confidence intervals.

\subsubsection{Penalized logistic regression}

Firth-penalized logistic regression models are implemented in \textsf{R} via the \texttt{logistf} package.\cite{logistf} Fitted values from the penalized model can be used directly within the $g$-computation framework in place of maximum-likelihood-based predictions. For inference, the penalized working model can be paired with either model-based or robust variance estimators, depending on the estimand of interest.

\newpage

\section{Simulation Study}

\subsection{Simulated response probabilities}

\begin{table}[h]
\centering
\setlength{\tabcolsep}{6pt}
\renewcommand{\arraystretch}{1.2}

\caption{Simulated conditional and marginal response probabilities and risk differences}
\label{tab:simRespProbs}

\begin{tabular}{c c c c c c c c c c}
\toprule
\multicolumn{3}{c}{} 
    & \multicolumn{2}{c}{$A=0$} 
    & \multicolumn{2}{c}{$A=1$} 
    & \multicolumn{3}{c}{} \\
\cmidrule(lr){4-5} \cmidrule(lr){6-7}

$\mathrm{RD}_A$ 
    & $\beta_1,\beta_2$ 
    & $ $ 
    & $X_2=0$ 
    & $X_2=1$ 
    & $X_2=0$ 
    & $X_2=1$ 
    & $E_{X=0}(Y)$ 
    & $E_{X=1}(Y)$ 
    & $\mathrm{RD}_X$ \\
\midrule

 &  & $X_1=0$ & 0.20 & 0.20 & 0.20 & 0.20 &  &  &  \\
\cmidrule(lr){3-7}
 & \multirow{-2}{*}{\(\log(1)\)} 
   & $X_1=1$ & 0.20 & 0.20 & 0.20 & 0.20 
   & \multirow{-2}{*}{0.20} 
   & \multirow{-2}{*}{0.20} 
   & \multirow{-2}{*}{0.00} \\
\cmidrule(lr){2-10}

 &  & $X_1=0$ & 0.14 & 0.20 & 0.14 & 0.20 &  &  & \\
\cmidrule(lr){3-7}
 & \multirow{-2}{*}{\(\log(1.5)\)} 
   & $X_1=1$ & 0.20 & 0.27 & 0.20 & 0.27 
   & \multirow{-2}{*}{0.17} 
   & \multirow{-2}{*}{0.23} 
   & \multirow{-2}{*}{0.06} \\
\cmidrule(lr){2-10}

 &  & $X_1=0$ & 0.07 & 0.17 & 0.07 & 0.17 &  &  & \\
\cmidrule(lr){3-7}
\multirow{-6}{*}{0} 
    & \multirow{-2}{*}{\(\log(3)\)} 
    & $X_1=1$ & 0.17 & 0.39 & 0.17 & 0.39 
    & \multirow{-2}{*}{0.12} 
    & \multirow{-2}{*}{0.28} 
    & \multirow{-2}{*}{0.16} \\
\midrule

 &  & $X_1=0$ & 0.20 & 0.20 & 0.35 & 0.35 &  &  & \\
\cmidrule(lr){3-7}
 & \multirow{-2}{*}{\(\log(1)\)} 
   & $X_1=1$ & 0.20 & 0.20 & 0.35 & 0.35 
   & \multirow{-2}{*}{0.28} 
   & \multirow{-2}{*}{0.28} 
   & \multirow{-2}{*}{0.00} \\
\cmidrule(lr){2-10}

 &  & $X_1=0$ & 0.14 & 0.20 & 0.26 & 0.35 & & & \\
\cmidrule(lr){3-7}
 & \multirow{-2}{*}{\(\log(1.5)\)} 
   & $X_1=1$ & 0.20 & 0.27 & 0.35 & 0.44 
   & \multirow{-2}{*}{0.24} 
   & \multirow{-2}{*}{0.31} 
   & \multirow{-2}{*}{0.08} \\
\cmidrule(lr){2-10}

 &  & $X_1=0$ & 0.07 & 0.17 & 0.14 & 0.33 & & & \\
\cmidrule(lr){3-7}
\multirow{-6}{*}{0.15} 
    & \multirow{-2}{*}{\(\log(3)\)} 
    & $X_1=1$ & 0.17 & 0.39 & 0.33 & 0.60 
    & \multirow{-2}{*}{0.18} 
    & \multirow{-2}{*}{0.37} 
    & \multirow{-2}{*}{0.19} \\
\midrule

 &  & $X_1=0$ & 0.20 & 0.20 & 0.50 & 0.50 &  &  & \\
\cmidrule(lr){3-7}
 & \multirow{-2}{*}{\(\log(1)\)} 
   & $X_1=1$ & 0.20 & 0.20 & 0.50 & 0.50 
   & \multirow{-2}{*}{0.35} 
   & \multirow{-2}{*}{0.35} 
   & \multirow{-2}{*}{0.00} \\
\cmidrule(lr){2-10}

 &  & $X_1=0$ & 0.14 & 0.20 & 0.40 & 0.50 & & & \\
\cmidrule(lr){3-7}
 & \multirow{-2}{*}{\(\log(1.5)\)} 
   & $X_1=1$ & 0.20 & 0.27 & 0.50 & 0.60 
   & \multirow{-2}{*}{0.31} 
   & \multirow{-2}{*}{0.39} 
   & \multirow{-2}{*}{0.08} \\
\cmidrule(lr){2-10}

 &  & $X_1=0$ & 0.07 & 0.17 & 0.25 & 0.50 & & & \\
\cmidrule(lr){3-7}
\multirow{-6}{*}{0.30} 
    & \multirow{-2}{*}{\(\log(3)\)} 
    & $X_1=1$ & 0.17 & 0.39 & 0.50 & 0.75 
    & \multirow{-2}{*}{0.25} 
    & \multirow{-2}{*}{0.45} 
    & \multirow{-2}{*}{0.21} \\
\bottomrule
\end{tabular}

\begin{tablenotes}
\footnotesize
\item Abbreviations. $\text{RD}_A, \text{RD}_X$ = marginal risk difference with respect to the treatment effect $A$ or any of the covariates $X_1, X_2$, respectively. $E_X(Y)$ = marginal response probability with respect to any of the covariates $X_1, X_2 = X$.
\end{tablenotes}
\end{table}

\FloatBarrier

\newpage

\subsection{Simulation results}

\begin{figure}[h]
    \centering
    \includegraphics[width=\textwidth]{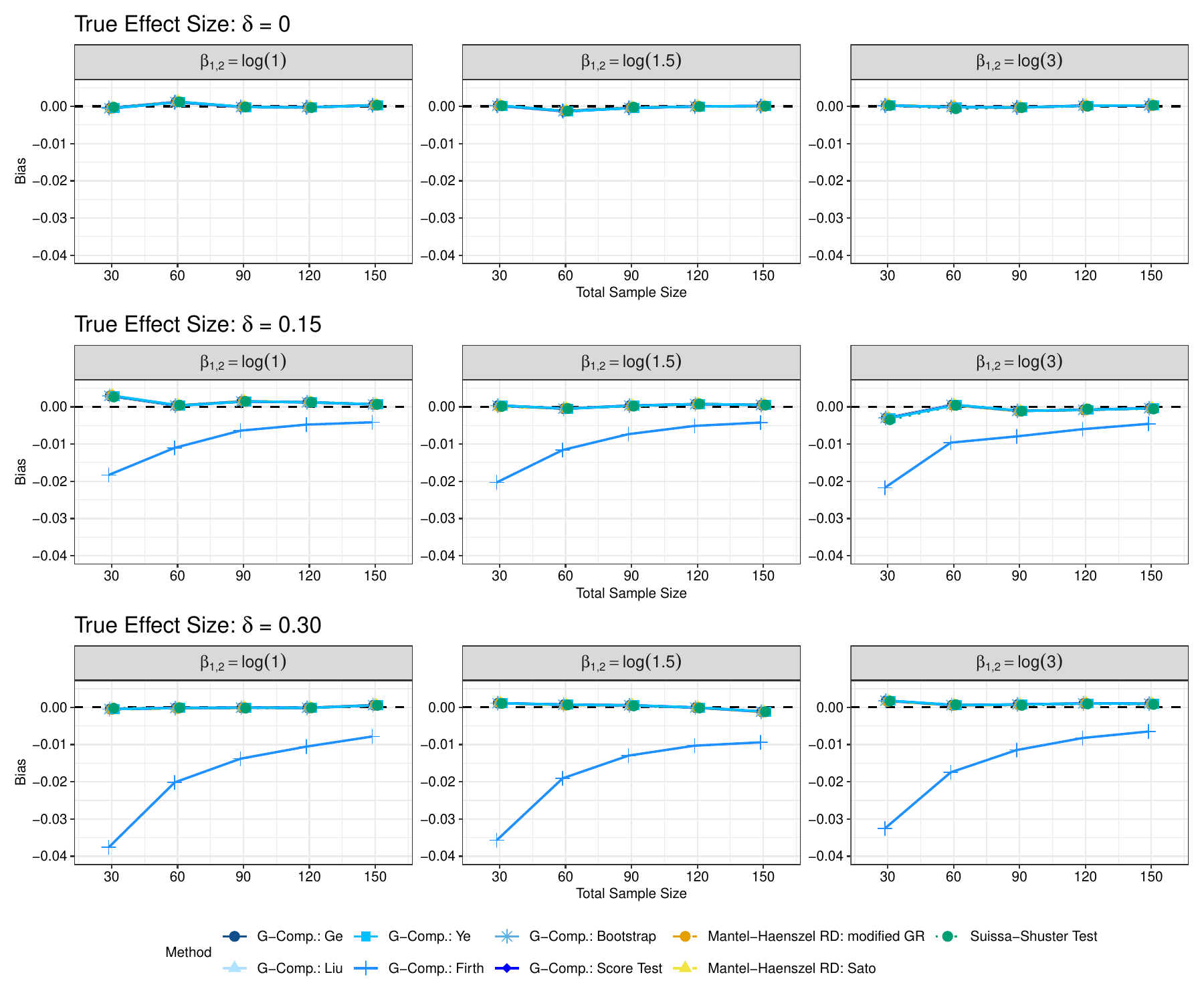}
    \caption{Estimation bias, i.e. the difference between the average estimated and the true risk difference, for the investigated statistical methods across simulation scenarios.}
    \label{fig:bias}
\end{figure}

\begin{figure}[h]
    \centering
    \includegraphics[width=\textwidth]{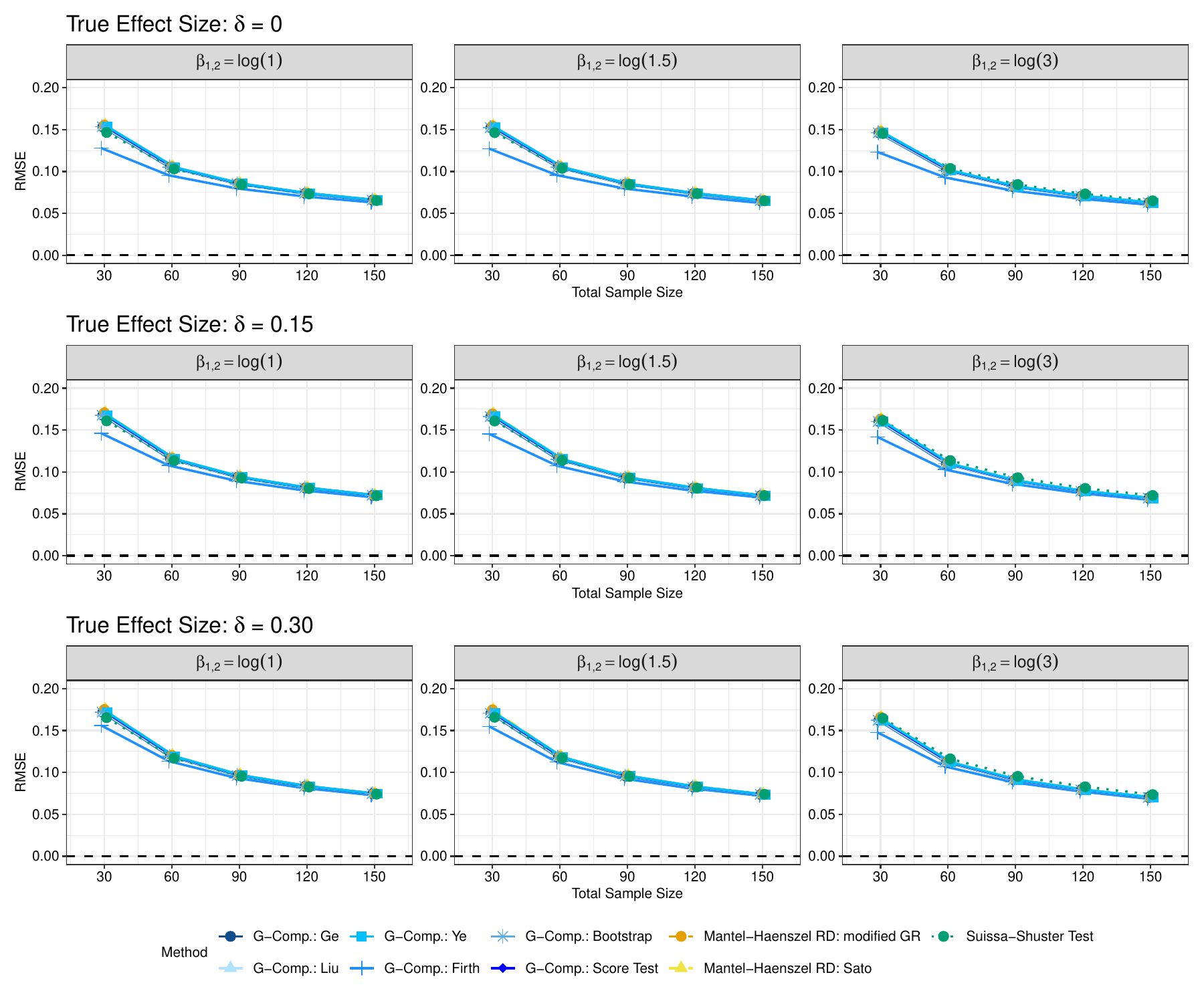}
    \caption{Root mean squared error for the investigated statistical methods across simulation scenarios.}
    \label{fig:rmse}
\end{figure}

\end{document}